\documentclass[aps,pre,
twocolumn,
superscriptaddress,
amsmath,amssymb,amsfonts]{revtex4-1}
\usepackage{graphicx}
\usepackage{hyperref}
\usepackage{color}
\begin{document}
\title{Thermodynamics of the Adsorption of Flexible Polymers on Nanowires}
\author{Thomas Vogel}
\email[Contact: ]{tvogel@lanl.gov}
\homepage[]{http://www.thomasvogel.name}
\affiliation{Theoretical Division (T-1), Los Alamos National Laboratory, Los Alamos, 
NM 87545, USA}
\author{Jonathan Gross}
\email[E-mail: ]{jonathan.gross@itp.uni-leipzig.de}
\affiliation{Institut f\"ur Theoretische Physik and Centre for Theoretical
Sciences (NTZ), Universit\"at Leipzig, Postfach 100\,920, D-04009 Leipzig,
Germany}
\affiliation{Soft Matter Systems Research Group, Center for Simulational 
Physics, The University of Georgia, Athens, GA 30602, USA}
\author{Michael Bachmann}
\email[E-mail: ]{bachmann@smsyslab.org}
\homepage[\\ Homepage: ]{http://www.smsyslab.org}
\affiliation{Soft Matter Systems Research Group, Center for Simulational 
Physics, The University of Georgia, Athens, GA 30602, USA}
\affiliation{Instituto de F\'{\i}sica,
Universidade Federal de Mato Grosso, 78060-900 Cuiab\'a (MT), Brazil}
\affiliation{Departamento de F\'{\i}sica,
Universidade Federal de Minas Gerais, 31270-901 Belo Horizonte (MG), Brazil}
\date{\today}
\begin{abstract}
Generalized-ensemble simulations enable the study of complex adsorption 
scenarios of a coarse-grained model polymer near an attractive nanostring, 
representing an ultrathin nanowire. We perform canonical and 
microcanonical statistical analyses to investigate structural transitions of 
the polymer and discuss their dependence on the temperature and on model 
parameters such as effective wire thickness and attraction strength. 
The result is a complete hyperphase diagram of the polymer phases, 
whose locations and stability are influenced by the effective material
properties 
of the nanowire and the strength of the thermal fluctuations. Major structural 
polymer phases in the adsorbed state include compact droplets attached to or 
wrapping around the wire, and tubelike conformations with triangular pattern 
that resemble ideal boron nanotubes. The classification of the 
transitions is performed by microcanonical inflection-point analysis. 
\end{abstract}
\maketitle
\section{Introduction}
\label{intro}
\enlargethispage{\baselineskip}
Although the recent advances in nanotechnology make it possible to study
smaller systems than ever before, the systematic understanding of miniature 
structures of macromolecules attached to inorganic adsorbents is an 
experimental challenge~\cite{whaley1,gray1}. The typical approaches of 
experimental verification of theoretical predictions and the theoretical 
understanding of experimental results is often limited to very specific 
questions~\cite{bgbgij1}. On the other hand, computer simulations are a 
comparatively inexpensive tool that allows for comprising studies, which can 
address generic problems as well. 
Pertaining to polymer adsorption at 
substrates, generic properties of this transition include common 
transition pathways from the desorbed into the different adsorption phases or 
vice versa (``wetting'', ``dewetting''), but also
transitions 
between structural phases under the influence of external factors such as
temperature 
and solvent quality. General material properties of the adsorbent, e.g., 
the propensity to bind macromolecules and the attraction (or repulsion)
range, affect the binding process as well.

There have been numerous computational studies of this complex behavior by 
means of simplified, coarse-grained models in the recent past. Yet, there are
still more questions than answers about the nature of the adsorption
phases that 
are difficult to approach. Strictly, most of these structural phases
are 
not phases in the thermodynamic sense, since the features of these phases and 
the transitions in between them are governed or at least substantially 
influenced by finite-size effects~\cite{mb1}. 
It has been a fairly profound finding that the thermodynamic behavior of
compact 
phases of finite polymers is rather irregular due to
discrete crystal symmetries and surface effects~\cite{vbj1,svbj1,sbj3}.

Most of the recent generic studies aiming at unraveling the structure
of the phase space and the adsorption dynamics have been performed for
polymers and proteins interacting with a perfectly flat substrate
with~\cite{bogner1,allen3,mjb2014} or without
pattern~\cite{vrbova,singh,causo1,prellberg1,bj4,linse1,paul1,mbj1,%
liang1,allen1,ivanov09jpcb,mjb2,ywli1}. Much less literature exists about
polymer adsorption at fluctuating surfaces such as
membranes~\cite{popova,kjb1}, because of the substantially increased
complexity of the problem. Flexible polymers can also optimally adapt to
curved surfaces such as cylindrical substrates~\cite{milchev1,srebnik,gvb1},
which causes the formation of polymer structures different from the adsorption
phases at flat substrates. 

Nanotube--polymer composites are a particularly interesting class of
hybrid systems with potential for nanotechnological applications.
Besides the individual mechanical and electronic properties of
nanotubes, in particular carbon
nanotubes~\cite{dressel01tap,jorio08buch}, hybrid nanotube--polymer
systems have an even broader spectrum of advanced applications in
technology and medicine. These features have been exploited, for
example, in photonics~\cite{hasan09am} and nanotubes coated with
conducting polymers were used to design biosensors~\cite{gao03ea}.
Potential medical applications include serum protein-coated gold
nano-rods for the selective targeting of cancer
cells~\cite{wang10nanolett,brewer05langmuir}. The understanding of the
polymer wetting behavior of nanotubes~\cite{tran08nanolett} is a key
to further developments in the technological preparation of such
hybrid systems.  The investigation of the interaction of
macromolecules with carbon nanotubes has also been the subject of
numerous computational
studies~\cite{wei06nanolett,tallury10jpcb,liu08jpcc,caddeo10jpcc,%
  wallace07nanolett,wallace10scale,angelik10lang}.

Recently, we investigated the extreme case of polymers adsorbed on
ultrathin nanowires and constructed a structural phase diagram based
on stable low-energy states~\cite{vb1}. This is particularly
interesting, because the nanowire geometry is topologically different
from a nanocylinder, which leads to a different monomer--substrate
interaction model. Therefore, to consider the nanowire as the limiting
case of a nanocylinder with zero radius is nontrivial.  Experimentally
resolved structures, for example for the spread of glycerol droplets
on carbon fibers and epoxy resin on aramid
filaments~\cite{wagner90jap} or liquid droplets at thin
cylinders~\cite{carroll86langmuir}, exhibit similarities to the
adsorbed, globular conformations we found in our simulations of a
simple coarse-grained polymer--nanowire model. The outer membrane of
certain bacteria possess an ordered, hexagonally packed surface layer
and although the units of the outer protein layer are not formed by
polymers, images of long tubular portions surrounded by hexagonally
packed units~\cite{glaeser79jur} resemble polymer-tube structures we
identified for strongly attractive nanowires~\cite{vb1}.  Potential
applications of hybrid polymer-nanowire assemblies include tube
systems for the transport of small molecules that can basically be
modeled in any desired shape.

In this paper, we extend our previous study of properties of lowest-energy
polymer adsorption phases~\cite{vb1} by
introducing the temperature as the external control parameter for thermal
fluctuations. By means of generalized-ensemble Monte
Carlo methods and by systematic variation of temperature and model parameters
such as effective wire thickness (volume exclusion) and attraction strength,
we investigate the thermodynamic properties of the structural phases of
adsorbed and desorbed polymers and the transitions between them. The
competition between energetic ordering and entropic diversity that causes
structural transitions will be investigated by the conventional canonical
statistical analysis of thermodynamic response quantities and more advanced
approaches such as the microcanonical inflection-point
analysis~\cite{mb1,infl1}. The latter method has proven to be
particularly useful, whenever finite-size effects dominate or substantially
influence structural transitions.

The paper is organized as follows. After the introduction of the
hybrid polymer--wire
model in Sect.~\ref{sec:model}, we briefly review
the conformational phase diagram of polymer ground-state structures
adsorbed to a nanowire in Sect.~\ref{sec:cpd}.
Results for the adsorption
thermodynamics of the polymer are analyzed and discussed in
Sect.~\ref{sec:canon_adsorb}. The microcanonical analysis of the adsorption
transition is performed in Sect.~\ref{sec:micro}.
Our main conclusions are summarized in
Sect.~\ref{sec:sum}.
\section{The polymer--wire model}
\label{sec:model}
We employ a generic, coarse-grained model for a self-interacting
homopolymer with $N$ monomers in the proximity of an attractive
one-dimensional stringlike substrate. The energy of a conformation
$\mathbf{X}$ of the polymer consists of contributions for the
interaction between nonbonded monomers, bending stiffness of the
polymer chain, and the interaction of each monomer with the nanowire:
\begin{eqnarray}
\label{eq:model}
E(\mathbf{X})=&&\sum_{i=1}^{N-2}\sum_{j=i+2}^{N}V_\mathrm{LJ}(r_{ij})+
\sum_{i=2}^{N-1}V_\mathrm{bend}(\cos\theta_i)\nonumber \\
&&+\sum_{i=1}^N V_\mathrm{string}(r_{\perp;i}).
\end{eqnarray}
The pairwise interaction between nonbonded
monomers is described by a standard Lennard-Jones potential,
\begin{equation}
\label{eq:LJ}
V_\mathrm{LJ}(r_{ij})=4\epsilon_\mathrm{m}
\left[ \left(\frac{
\sigma_\mathrm{m}}{r_{ij}}\right)^{12}-\left(\frac{\sigma_\mathrm{m}}{r_{ij}}
\right)^{6}\right],
\end{equation}
where $r_{ij}$ is the distance between monomers $i$ and $j$. 
The potential minimum
$V_\mathrm{LJ}(r_{ij}^\mathrm{min})=-\epsilon_\mathrm{m}$ is located at
$r_{ij}^\mathrm{min}=2^{1/6}\sigma_\mathrm{m}$.
In our simulations, we used energy and
length units for which $\epsilon_\mathrm{m}=1$ and $\sigma_\mathrm{m}=1$,
respectively. 
Bonds between monomers adjacent along the chain are stiff and of unit length,
$r_{i\,i+1}=1$. Covalent bond vectors connected to the $i$th
monomer form a bending angle $\theta_i$. The bending energy
\begin{equation}
\label{eq:bend}
V_\mathrm{bend}(\cos\theta_i)=\kappa (1-\cos\theta_i)
\end{equation}
is a remnant of the AB model, which has initially been introduced as a
coarse-grained hydrophobic--polar peptide model~\cite{still1} that has proven
to be useful for the qualitative description of generic features of
conformational transitions associated with protein folding~\cite{baj1} and
aggregation processes~\cite{jbj1}. In this model, the bending stiffness is set
to $\kappa=1/4$. 
The homopolymer version of the model has also been used in recent adsorption
studies of polymers~\cite{mjb2014,mbj1,mjb2,vb1,vb2}. 

For the determination of the string potential, we consider
a monomer interacting with each infinitesimal element of the string, which
is supposed to extend infinitely into $\pm z$ direction. We assume that this
interaction is of van der Waals type and
can be described by a Lennard-Jones potential. By
using the cylindrical coordinate representation, the total potential
of a monomer $i$ in the field of the string is then given by~\cite{vb1,vb2}
\begin{eqnarray}
\label{eq:string}
V_\mathrm{string}(r_{\perp;i})&=& 4\eta_\mathrm{f}\epsilon_\mathrm{f}\int
\limits_{-\infty}^\infty
\mathrm{d}z\,\left[\frac{\sigma_\mathrm{f}^{12}}{(r_{\perp;i}^2+z^2)^6}
-\frac{\sigma_\mathrm{f}^6}{(r_{\perp;i}^2+z^2)^3}\right]\nonumber \\
&=& \pi \,
\eta_\mathrm{f}\epsilon_\mathrm{f}\left(\frac{63}{64}
\frac{\sigma_\mathrm{f}^{12}}{r_{\perp;i}^{11}}-\frac{3}{2}
\frac{\sigma_\mathrm{f}^{6}}{r_{\perp;i}^5}
\right),
\end{eqnarray}
where we have introduced the perpendicular distance $r_{\perp;i}$ of the
monomer from the string axis and the van der Waals radius or effective
``thickness'' of the string $\sigma_\mathrm{f}$, which is related to the
potential minimum distance $r_\perp^\mathrm{min}$ via:
\begin{equation}
\label{eq:perpmin}
r_\perp^\mathrm{min}=\left(\frac{693}{480}\right)^{1/6}\sigma_\mathrm{f}
\approx 1.06\sigma_\mathrm{f}.
\end{equation}
The string ``charge'' density
$\eta_\mathrm{f}$ compensates for the additional dimension by the extra 
$\sigma_\mathrm{f}$ factor that remains after integration over the standard
form of the Lennard-Jones potential. It can be conveniently used to set the
energy scale of the potential. We adjust it in such a way that the minimum
monomer--string energy is
$V_\mathrm{string}(r_\perp^\mathrm{min})=-\epsilon_\mathrm{f}$, independently
of $\sigma_\mathrm{f}$. In this case,
$\eta_\mathrm{f}\approx 0.528/\sigma_\mathrm{f}$.

\begin{figure}
\includegraphics[width=.98\columnwidth]{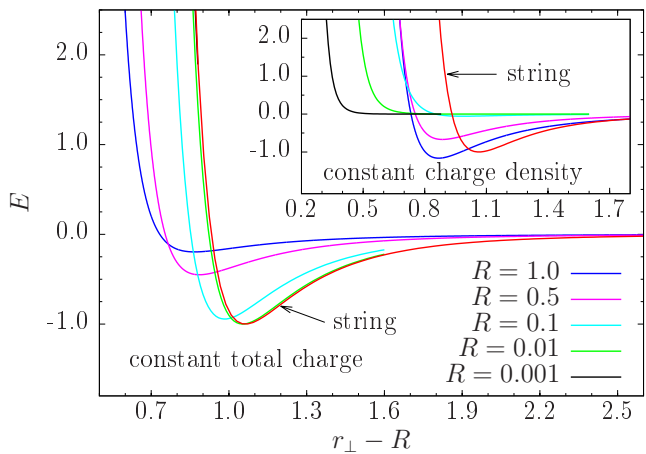}
\caption{\label{fig:comp}Cylinder potential [Eq.~(\ref{eq:cyl})] for
  different radii $R\in (0\ldots1]$ and the string potential
  [Eq.~(\ref{eq:string})] for
  $\sigma_\mathrm{f}=\epsilon_\mathrm{f}=1$ for constant \textit{total
    ``charge''}, i.e., $\eta_\mathrm{cyl}=1/\pi R^2$ (main figure) and
  for constant \textit{``charge'' density} $\eta_\mathrm{cyl}=1$
  (inset). The inset figure corresponds to Fig.~4 in
  Ref.~\cite{milchev1}.}
\end{figure}
It is instructive to view the string potential~(\ref{eq:string}) from a
different perspective, namely as the $R\to 0$ limit of the corresponding
potential of an attractive cylinder. This is nontrivial, because of the
different topological
dimensions of string and cylinder.
The cylinder--monomer potential can be written as~\cite{milchev1}:
\begin{eqnarray}
&&V_\mathrm{cyl}(R,r_{\perp;i})=\pi\epsilon_\mathrm{f}
\eta_\mathrm{cyl}
\int_0^{2\pi}\mathrm{d}\phi\int_0^R\rho\,\mathrm{d}\rho\nonumber\\
&&\hspace{12mm}\times\left(\frac{63}{64}\frac{\sigma_\mathrm{f}^{12}}
{{r'}^{11}(r_{\perp;i},\rho,\phi)}-\frac{3}{2}\frac{\sigma_\mathrm{f}^{6}}
{{r'}^5(r_{\perp;i},\rho,\phi)}\right),
\label{eq:cyl}
\end{eqnarray}
where the integration in $z$ direction has already
been performed. The distance between the $i$th monomer and any point in the
cylinder disk in-plane
with the monomer is given
by $r^\prime=(r_{\perp;i}^2 +\rho^2-2\rho\,
r_{\perp;i}\,\cos\phi)^{1/2}$; $\rho$ and $\phi$ are the
polar coordinates in the plane of cylinder disk and monomer. Compared to
Ref.~\cite{milchev1}, we have introduced in the potential~(\ref{eq:cyl}) the
\emph{volume ``charge'' density} $\eta_\mathrm{cyl}$ of the cylinder. If this
density is considered a constant and is thus independent of $R$ (as in
Ref.~\cite{milchev1}), the string potential~(\ref{eq:string}) \emph{cannot}
be obtained as the $R\to 0$ limit of the cylinder potential. Only if the
\emph{total ``charge''} is fixed, i.e., $\eta_\mathrm{cyl}$ scales with
the inverse area of the cylinder disk $1/\pi R^2$, then
\begin{equation}
V_\mathrm{string}(r_{\perp;i})=\lim_{R\to 0}V_\mathrm{cyl}(R,r_{\perp;i}).
\end{equation}
For both cases, the $R\to 0$ limits of the cylinder potential are plotted in
Fig.~\ref{fig:comp}.
The systematic discussion of polymer adsorption at cylindrical substrates
will be done elsewhere~\cite{gvb1}.

In our model of the polymer--wire system, the polymer is not grafted to
the string and can move freely. The
string, pointing into $z$
direction, is located in the center of the periodic simulation box with edge
lengths $2N$ in $x$ and $y$ directions to prevent the polymer from escaping.

In our simulations, we performed different runs using
multicanonical~\cite{muca,janke98physA,mb2} and Wang-Landau
sampling~\cite{wl_prle,zhou05pre} and found the results to be
coherent.  Conformational updates included local crankshaft and
slithering-snake moves, as well as global spherical-cap~\cite{baj1}
and translation moves~\cite{vb2}.  The crankshaft move corresponds to
the rotation of a single monomer about the axis given by the vector
between its nearest neighbors and changes the conformation locally.
For the slithering-snake update, a monomer is cut at one end of the
chain and pasted at the other end, inverting the bond vector. This
move proves useful to avoid trapping in very dense conformations. The
spherical-cap update consists of a pivot rotation of a single bond and
the subsequent shift of the monomers at either end of the polymer,
keeping all bond lengths fixed. This update allows for larger steps in
the conformational space.  Finally, the global translation update
allows for a displacement of the entire polymer chain relative to the
string. Update types were chosen randomly with equal weight, which
allowed for the efficient sampling of the conformational space.

The chain lengths studied varied from $N=30$ to $N=100$ for the
analysis of the thermodynamic adsorption behavior and up to $N=200$ for the
investigation of ground-state properties.
\section{Lowest-energy conformations of adsorbed polymers}
\label{sec:cpd}
The investigation of conformational properties of lowest-energy states
is generally beneficial for the identification of compact crystalline
or amorphous structural phases. The major ground-state morphologies of
the polymer interacting with an attractive nanowire were found to be
string-attached globular droplets and tubelike barrels wrapped around
the string~\cite{vb1}. In the first case, depending on the attraction
strength and effective thickness of the wire, the string penetrates
the globule (Gi: globular, included) or it is only loosely attached to
it (Ge: globular, excluded).
\begin{figure}
\includegraphics[width=\columnwidth]{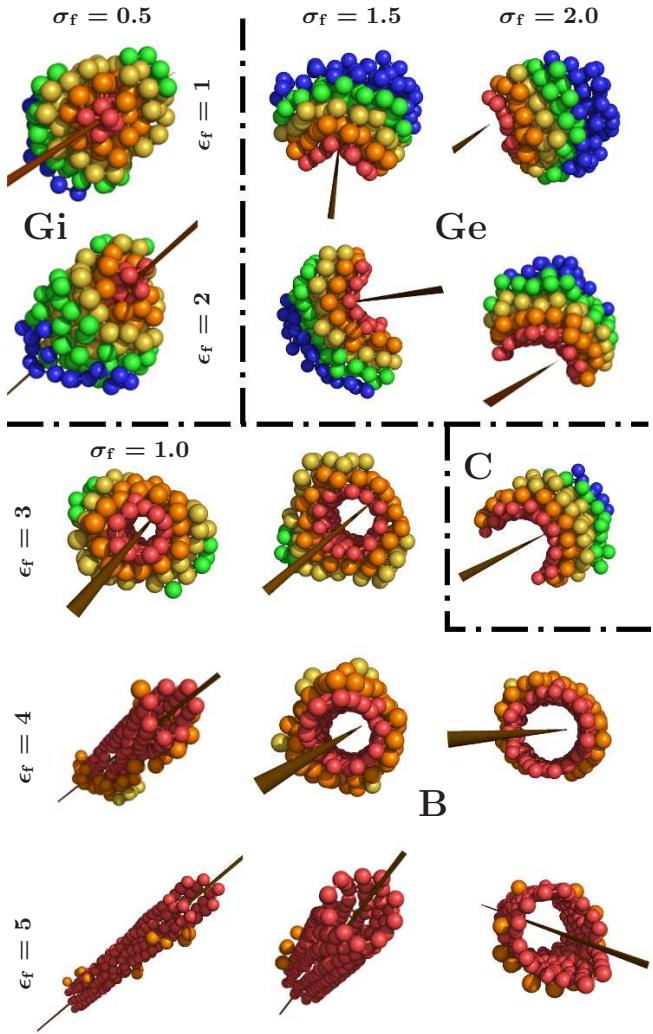}
\caption{\label{fig:l200}Typical low-energy conformations of a polymer with
$N=200$
monomers adsorbed at a string with effective thicknesses
$\sigma_\mathrm{f}$-values $0.5$ and $1$, respectively (first
column),
$1.5$ (second) and $2$ (third) for attraction strengths
$\epsilon_\mathrm{f}=1,2,3,4$, and $5$ (top to bottom). Monomers are
marked with different shades
according to their distance from the string.\vspace*{-\baselineskip}}
\end{figure}

Most interesting is the formation of tubelike, or barrellike (B), 
polymer structures, which resemble nanotube morphologies~\cite{vmab1}. This 
occurs if the string
field is more attractive for most monomers than the compact crystalline
assembly and the effectively excluded string diameter is small enough.

Topological
transitions, common to all adsorption problems and typically accompanied
by layering effects, can also be observed in the barrel phase. At very low
$\sigma_\mathrm{f}$ values, where
monomers are attached so close to the string that they can feel the attractive
field of monomers of the opposite side of the string as well, the
polymer ``tube'' looks rather like a one-dimensional chain with zig-zag (or
sawtooth) pattern. In the other extreme case, if the effective string
thickness is
too large, additional monomers will attach to the
two-dimensional monolayer
tube, essentially forming a new layer. Three-dimensional
assemblies with two or more layers are also known from atomic
nanotubes~\cite{wang1}. 

For comparatively large string thickness and sufficiently high
adsorption strength, the polymer wraps only partially around the
string. This structural phase is very typical for polymer adsorption
transitions at cylinders~\cite{milchev1,gvb1} and has been called
``clamshell'' phase (C).

Figure~\ref{fig:l200} shows typical conformations in these different
structural low-energy phases for a polymer with 200 monomers at
various values of string parameters $\sigma_\mathrm{f}$ and
$\epsilon_\mathrm{f}$, supporting our general findings for other chain
lengths~\cite{vb1}.  Since the monomer--string interaction is
attractive, the polymer is adsorbed to the string in all low-energy
phases.

In the
following, we will investigate the influence of thermal fluctuations upon
the the formation of compact adsorbed polymer structures and the 
thermodynamic properties of the adsorption
transition.

\section{Thermodynamics of polymer adsorption at stringlike substrates}
\label{sec:canon_adsorb}
For the understanding of the thermodynamic behavior of the system, it
is useful to perform canonical and microcanonical analyses of various
thermodynamic quantities with the goal of identifying generic features
of polymer adsorption at a stringlike nanowire for various
parametrizations in the space of effective van der Waals radius
$\sigma_\mathrm{f}$ of the wire and its attractive adsorption strength
$\epsilon_\mathrm{f}$.

\begin{figure}
\includegraphics[width=.95\columnwidth]{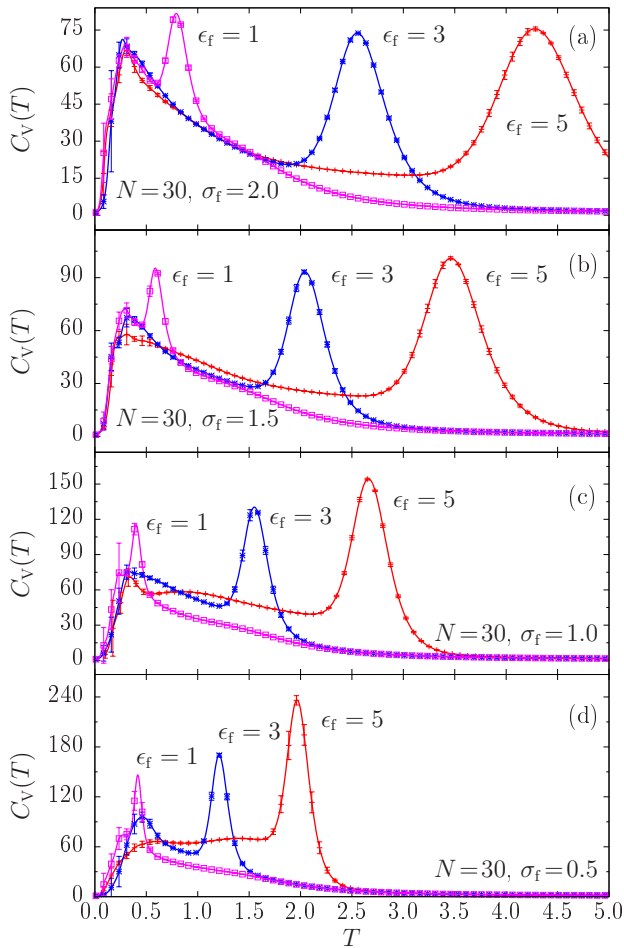}
\caption{\label{fig:cv}Temperature dependence of the heat
capacities of a 30-mer for different
string attraction
strengths ($\epsilon_\mathrm{f}=1$, $3$, and $5$, respectively) and
decreasing van der Waals radii: (a) $\sigma_\mathrm{f}=2.0$, (b)
$1.5$, (c) $1.0$, and
(d) $0.5$. Note the different ordinate scales. Error bars were estimated using
the jackknife method~\cite{mb1}.}
\end{figure}

Figure~\ref{fig:cv} shows heat-capacity curves $C_V(T)$ for a polymer
consisting of $N=30$ monomers interacting with various nanowires,
distinguished by their different material parameters
$\sigma_\mathrm{f}$ and $\epsilon_\mathrm{f}$. Most pronounced in all
cases is the significant peak at high temperatures, which corresponds
to the adsorption transition. For fixed values of $\sigma_\mathrm{f}$,
the peak position, which we denote by $T_\mathrm{ads}^\mathrm{can}$ in
the following, scales almost perfectly linearly with the wire
attraction strength, $T_\mathrm{ads}^\mathrm{can}\propto
\epsilon_\mathrm{f}$. This has already been found in adsorption
studies at planar substrates~\cite{bj4,mbj1} and can intuitively be
understood by the larger thermal energy $\sim k_\mathrm{B}T$ needed to
release the polymer off the substrate at larger adsorption strengths.
However, it is worth mentioning as well that the proportionality
factor depends nonlinearly on the effective string thickness
$\sigma_\mathrm{f}$.

\begin{figure}
\includegraphics[width=\columnwidth]{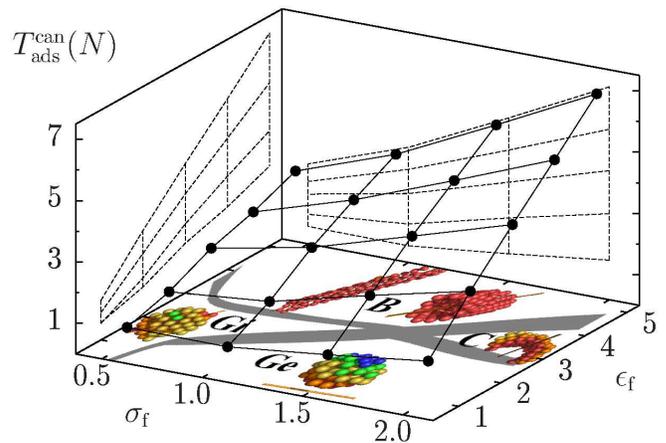}
\caption{\label{fig:cvpeaks}Adsorption temperatures for different values of
$\epsilon_\mathrm{f}$
and effective wire thickness $\sigma_\mathrm{f}$ for a
polymer of length $N=100$. The structural phase diagram for
ground-states from Ref.~\cite{vb1} is projected upon
the $T=0$ plane.}
\end{figure}

This can also be seen in Fig.~\ref{fig:cvpeaks}, where we have plotted
the adsorption temperature as a function of $\sigma_\mathrm{f}$ and
$\epsilon_\mathrm{f}$ for a 100-mer. We have also included the
zero-temperature structural phase diagram from Ref.~\cite{vb1} into
the $T=0$ plane to get an impression of how the compact adsorption
phases at given $\sigma_\mathrm{f}$ and $\epsilon_\mathrm{f}$ values
look like. Not surprisingly, the adsorption transition temperatures
depend on $N$ and are different for the 100-mer, compared to the
30-mer. However, qualitatively the generic results are comparable.

After passing the adsorption transition, for temperatures
$T_\mathrm{cryst}^\mathrm{can}<T<T_\mathrm{ads}^\mathrm{can}$,
dominant polymer
conformations are
either disordered and expanded or locally ordered and globular. For the
adsorption of the 30-mer at the nanowire, these phases are difficult to
distinguish. More striking is the crystallization transition at
$T_\mathrm{cryst}^\mathrm{can}$ into the compact
adsorption phases that have been described qualitatively in
Section~\ref{sec:cpd}. As it is obvious from Fig.~\ref{fig:cv}, the
low-temperature peak positions in the specific heat curves near
$T_\mathrm{cryst}^\mathrm{can}$, indicating this transition, do not vary much
and, therefore, hardly depend on the material parameters of the nanowire,
$\epsilon_\mathrm{f}$ and $\sigma_\mathrm{f}$. Effectively, during this
transition, the number of contacts between monomers and wire does not change
much and the way in which the polymer binds to the wire has already been
established in the ``liquid'' (expanded or globular) phase. Note that, for
strong attraction (large $\epsilon_\mathrm{f}$ values), the crystallization
transition becomes weaker, the smaller the effective wire thickness is. This
confirms the previous statement that the basic, tubelike structures have
already formed in the ``liquid'' phase, with most monomers already in contact
to the wire.

If the wire attraction and thickness are comparatively small
($\epsilon_\mathrm{f}<2$ and $\sigma_\mathrm{f}<1$), the chain behaves
basically like a free, flexible polymer, with minimal recognition of the wire,
which is enclosed inside the polymer conformation for energetic reasons. In
this case, the adsorbed liquid phase is preempted by the compact, crystalline
phase. Adsorption and crystallization merge. Noteworthy are the very weak
peaks in Fig.~\ref{fig:cv} at about
$T_\Theta^\mathrm{can}\approx 1.5$, which indicate the $\Theta$ transition
of the polymer chain. The weakness of the $\Theta$ transition for very short
flexible and semiflexible chains has already been observed
previously~\cite{vbj1,svbj1,sbj3}. 

\begin{figure}
\includegraphics[width=\columnwidth]{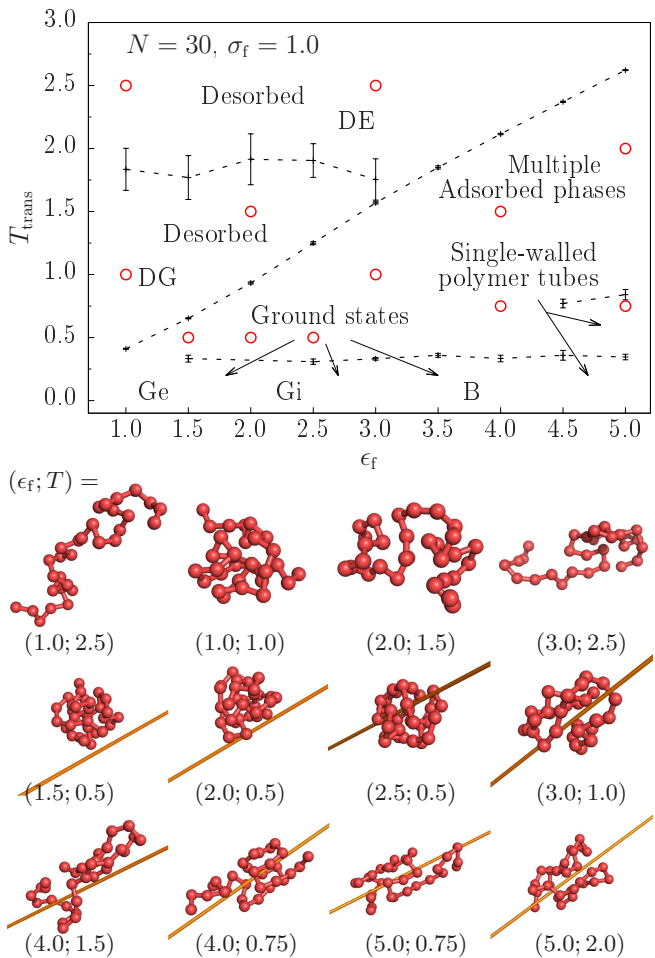}
\caption{Example of the structural phase diagram and representative
conformations in the different conformational phases for $N=30$ at wire
thickness
$\sigma_\mathrm{f}=1.0$. Transition temperatures $T_\mathrm{trans}$ were
identified from peaks in the specific heat. The low-energy ``clamshell'' phase
C does not exist for $\sigma_\mathrm{f}=1.0$.}
\label{fig:thermo_complete}
\end{figure}

Figure~\ref{fig:thermo_complete} shows an example for a complete
structural phase diagram at $\sigma_\mathrm{f}=1.0$ for $N=30$.
Representative conformations in the different phases at parameter
values indicated by the circles in the phase diagram are also
depicted. The phase diagram prominently features the almost linear
adsorption transition line between desorbed and adsorbed phases.
Transitions from the desorbed extended (DE) and desorbed globular (DG)
phases, which are separated by the well-known $\Theta$ transition
line, into adsorbed phases lead to a variety of structures. However,
given the rather small size of the polymer, these conformations can
hardly be distinguished. Although we expect that the dominant DE
structures will also adsorb as extended conformations and, similarly,
DG structures cross over into adsorbed globules, a quantitative
description is more difficult than in the adsorption scenarios at
planar substrates~\cite{bj4,mbj1}.  A significant signal for the
separation of extended and globular in the adsorbed regime could not
be identified. The ``crystallization'' transition into compact phases
dominated by the ground-state structures, however, is clearly visible.
In complete analogy to the previous analysis of ground-state
structures for $\sigma_\mathrm{f}=1$ in the case of the
100-mer~\cite{vb1} (see also the $\sigma_\mathrm{f}=1$ cross section
in the projection in Fig.~\ref{fig:cvpeaks}), we clearly identify
compact globular droplets with the string excluded (Ge) at rather
small values of the string attraction strength $\epsilon_\mathrm{f}$.
With $\epsilon_\mathrm{f}$ increasing, the string becomes included
into the globule (Gi). Only for sufficiently large string attraction
($\epsilon_\mathrm{f}>3$), tubelike ``barrel'' (B) structures form.
Further increasing $\epsilon_\mathrm{f}$, these become dominated by
monolayer (or single-walled) polymer tubes. The detailed discussion of
the ground-state dominated compact phases can be found in
Ref.~\cite{vb1}.
\section{The adsorption transition from the microcanonical perspective}
\label{sec:micro}
\begin{figure}
\includegraphics[width=.9\columnwidth]{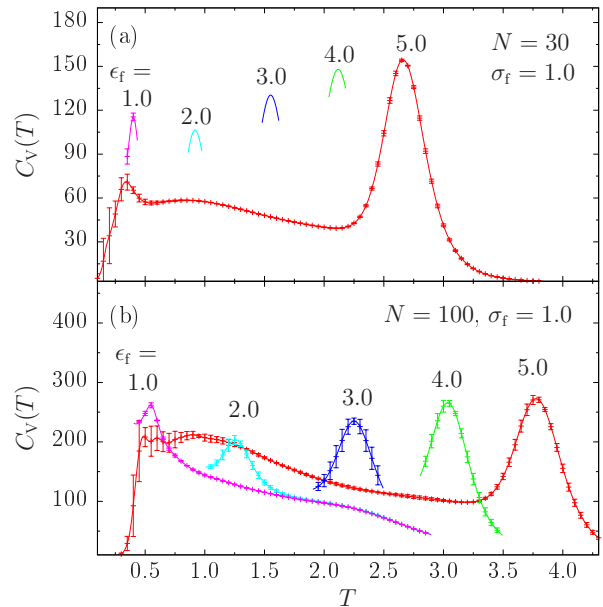}
\caption{
\label{fig:thermoX}
Adsorption transition signals in the heat-capacity curves of
(a) a 30-mer and (b) a 100-mer for
different values of the wire attraction strength $\sigma_\mathrm{f}$ and
fixed effective thickness $\sigma_\mathrm{f}=1.0$. For visualization
purposes, only the adsorption peaks are shown for $\epsilon_\mathrm{f}<5.0$.
}
\end{figure}
The canonical statistical analysis of different response quantities,
including the specific heat, as functions of the canonical temperature is
typically ambiguous. Although signals such as extremal points and inflection
points are helpful indicators of thermal activity, they do not allow for a
unique identification of transition points~\cite{mb1,baj1,bj1}. 
Figure~\ref{fig:thermoX} shows the adsorption peaks in the heat-capacity
curves for various $\varepsilon_\mathrm{f}$ values and wire thickness
$\sigma_\mathrm{f}=1.0$ for a 30-mer and a 100-mer.

An alternative
method that has proven to be extremely useful is based on the fact that
temperature can also be defined as a system property. This method is called
microcanonical inflection-point analysis~\cite{infl1} and rests upon the
shoulders of microcanonical statistical analysis~\cite{gross1}.\vadjust{\break}

\begin{figure}
\includegraphics[width=\columnwidth]{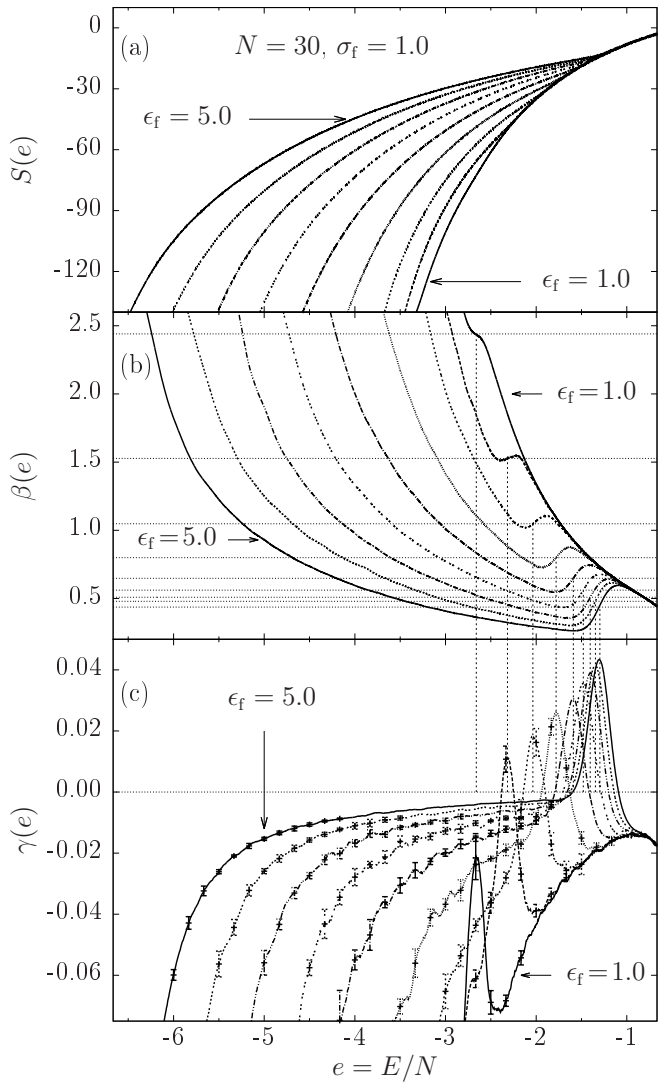}
\caption{Microcanonical analysis of structural transitions of the
polymer with $N=30$ monomers
interacting with wires of effective thickness $\sigma_\mathrm{f}=1$ and
attraction strengths $\epsilon_\mathrm{f}=1.0,1.5,2.0,\ldots,5.0$.
(a) Microcanonical entropy $S(e)$, (b) inverse microcanonical temperature
$\beta(e)=(1/N)(dS(e)/de)$, and (c) inflection-point indicator
$\gamma(e)=(1/N)(d\beta(e)/de)$.
}
\label{fig:micro}
\end{figure}

\begin{figure}
\includegraphics[width=\columnwidth]{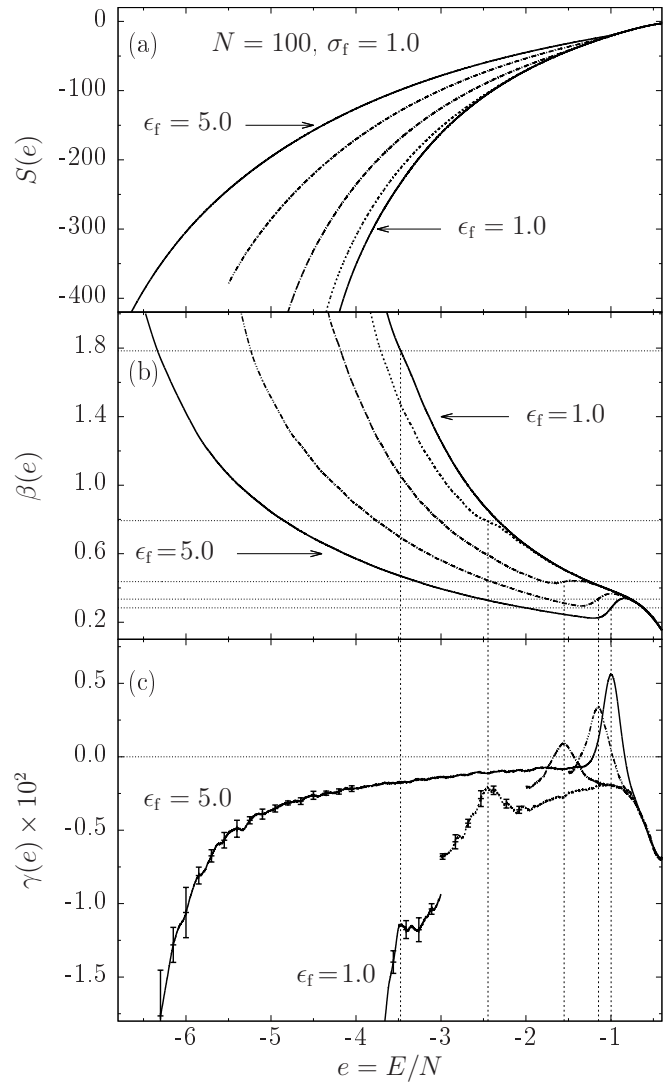}
\caption{Same as Fig.~\ref{fig:micro}, but for $N=100$ and
attraction strengths $\epsilon_\mathrm{f}=1.0,2.0,\ldots,5.0$.
}
\label{fig:micro100}
\end{figure}

The central quantity is the density of states $g(E')=\int {\cal
  D}X\,\delta(E'-E(\textbf{X}))$, which sets entropy and energy into
relation with each other: $S(E)=k_\mathrm{B}\ln\, g(E)$. It is easy to
show that microcanonical equilibrium between two systems is achieved
if $\partial S_1(E_1)/\partial E_1=\partial S_2(E_2)/\partial E_2$.
With this, the inverse microcanonical temperature can be defined as
$\beta(E)=\partial S(E)/\partial E$. Figures~\ref{fig:micro}(a)
and~(b) show for the 30-mer the microcanonical entropy and inverse
temperature curves, $S(E)$ and $\beta(E)$, respectively, for wire
material parameters $\varepsilon_\mathrm{f}=1.0,1.5,2.0,\dots,5.0$ and
$\sigma_\mathrm{f}=1.0$. For comparison, the same plots are shown in
Fig.~\ref{fig:micro100} for a 100-mer.

The $\beta(E)$ curves in Figs.~\ref{fig:micro}(b)
and~\ref{fig:micro100}(b) exhibit a characteristic behavior that help
locate transition points uniquely. Specific
monotonic properties can even be used to introduce a systematic classification
scheme. In microcanonical inflection-point analysis~\cite{infl1}, the
transition points
are defined by means of the derivative $\gamma(E)=d\beta(E)/dE$ in the
following way. If $\gamma(E)$ exhibits a local maximum and the extremal value
is positive, $\gamma_\mathrm{max}(E_\mathrm{tr})>0$, then the associated
inflection point of the microcanonical temperature at $E_\mathrm{tr}$ must lie
in a low-sensitivity region, in which an energetic increase has a minimal
effect on the change of temperature within a certain energetic range. For a
finite system, this entails
the ``backbending'' of the temperature curve, which levels out in the
thermodynamic limit
and renders the Maxwell construction exact. The energy that
spans the transition range is associated with a nonzero latent heat $\Delta
Q>0$, which is why this transition is referred to as a \emph{first-order
transition}.

Consequently, a negative local maximum
$\gamma_\mathrm{max}(E_\mathrm{tr})<0$,
which also signals a low-sensitivity region of the $\beta(E)$ curve, marks a
\emph{second-order transition}, because it corresponds to a transition
\emph{point} in energy space and, therefore, $\Delta Q=0$.
Note that according to these definitions the order of transition
depends on the system size. It is not unusual that a transition with
first-order signature in a finite system turns into a continuous,
second-order phase transition in the thermodynamic limit. Not
surprisingly, the order of a transition is also sensitive to changes of
system (or model) parameters.

Most strikingly, as can be seen in Figs.~\ref{fig:micro}(b) and~(c)
for the 30-mer, the
adsorption transition that we classify as a second-order transition at
$\varepsilon_\mathrm{f}=1.0$, turns into first order for larger attraction
strengths. The first-order characteristic becomes stronger with increasing
$\varepsilon_\mathrm{f}$ values. The horizontal lines in
Fig.~\ref{fig:micro}(b) represent the
(inverse) transition temperatures identified from the corresponding
inflection
points. They coincide very nicely with the peak values of the
specific heat curves shown in Fig.~\ref{fig:thermoX}. The scenario
is similar for the 100-mer. The inverse microcanonical temperatures,
plotted in Fig.~\ref{fig:micro100}(b) for various
$\varepsilon_\mathrm{f}$ values, exhibit features of a second-order
transition for $\varepsilon_\mathrm{f}<3.0$, before turning to first order
at higher $\varepsilon_\mathrm{f}$ values.
This is very similar to
the previously discussed case of polymer adsorption at flat
substrates~\cite{mjb2}. 

The reason for the increasing entropic suppression of
states in the phase separation regime is that less states are available for
each energetic state in the linearly increasing energy interval (the
energy scale grows like $E\propto \varepsilon_\mathrm{f}$). Since the
desorption phase is almost entirely
governed by translational entropy (in a constant energy interval above the
adsorption transition), and the energetically controlled adsorption states are
shifted to lower energies (below the adsorption transition) with
$\varepsilon_\mathrm{f}$ increasing, the energetic space in between is
entropically ``emptied'' (note that the total number of states is independent
of $\varepsilon_\mathrm{f}$). For the same reason, the slope of the
double-tangents (Gibbs construction) at the respective transition points
decreases for larger values of $\varepsilon_\mathrm{f}$, which
effectively shifts the adsorption transition toward higher temperatures,
which is consistent with the canonical analysis of Fig.~\ref{fig:thermoX}.
Since the energy scale of adsorption is linear, the adsorption transition
temperature scales almost linearly as well over a large range of the
adsorption strength parameter, $T_\mathrm{ads}\sim\varepsilon_\mathrm{f}$,
which is a common feature of all minimally (additively) coupled adsorption
models~\cite{bj4,mbj1,mjb2}.

The increased $\varepsilon_\mathrm{f}$ range of
second-order behavior of the 100-mer, compared to the 30-mer, is an
empirical indication that the adsorption transition of a polymer at a
linelike substrate is a
second-order phase transition
in the thermodynamic limit. This is consistent with the general
assumption that adsorption transitions of infinitely large systems are
continuous~\cite{kremer1}. This means that
the width of the phase-separation regime, which is nonzero in the
finite system, shrinks with increasing number
of monomers and the per-particle latent heat diminishes for $N\to
\infty$~\cite{mjb2}.
\section{Summary}
\label{sec:sum}
We have investigated the thermodynamic behavior of structural phases
for a hybrid model of a polymer interacting with stringlike substrate
that resembles a~nano\-wire. This study extends the previous study of
ground-state properties of the same system~\cite{vb1} by analyzing the
influence of thermal fluctuations upon the adsorption transition and
the formation of compact adsorption phases. Results were obtained by
means of advanced generalized-ensemble Monte Carlo simulations. We
identified all major structural thermodynamic phases from indications
of enhanced thermal activity in the specific heat curves and
constructed the structural hyperphase diagram, parametrized by
temperature and string attraction strength. As expected, for
increasing attraction strength, the low-temperature structures change
from compact crystalline shapes that attach to the string or include
it to barrellike conformations that resemble single-walled nanotubes.
The transition into less ordered adsorption phases turns out to be
independent of the adsorption strength, i.e., the transition
temperature is virtually constant. In contrast, the
adsorption/desorption transition temperature scales linearly with the
substrate attraction strength. A microcanonical inflection-point
analysis revealed that the adsorption transition possesses an
increasingly strong first-order characteristic, the larger the
attraction strength is. However, it is important to note that this
first-order behavior is a finite-system feature that is supposed to
disappear in the thermodynamic limit. It is only due to an effective,
relative suppression of microstates in the phase-separation transition
region, which is a finite-size effect. This does not contradict the
common view that the adsorption transition is a second-order phase
transition in the thermodynamic limit.

The results presented in this paper provide general background knowledge
needed for systematic approaches to the technological design of functional
materials used for the transport of molecules (based on polymer tubes) and
also give insight into binding options of biomacromolecules at nanofibers
(such as actin filaments) in cells.
\begin{acknowledgments}
  This work has been supported partially by the NSF under Grant No.\
  DMR-1207437, by CNPq (National Council for Scientific and
  Technological Development, Brazil) under Grant No.\ 402091/2012-4,
  and by the DFG (German Research Foundation) under Grant No.\
  SFB/TRR~102 (Project B04).  The authors also acknowledge support by
  the J\"u\-lich/Aachen/Haifa Umbrella program under Grants No.~SIM6
  and No.~HPC\_2. Computer time was provided by the
  For\-schungs\-zentrum J\"ulich under Project No.~jiff39 and
  No.~jiff43, and by the Georgia Advanced Computing Resource Center
  (GACRC) at the University of Georgia.  Assigned: LA-UR-14-27917.
\end{acknowledgments}

\begin{thebibliography}{99}%
%
\bibitem{whaley1}
S.\ R.\ Whaley, D.\ S.\ English, E.\ L.\ Hu, P.\ F.\ Barbara, A.\ M.\ Belcher,
Nature \textbf{405}, 665 (2000).
%
\bibitem{gray1}
J.\ J.\ Gray, Curr.\ Opin.\ Struct.\ Biol.\ \textbf{14}, 110 (2004).
%
\bibitem{bgbgij1} M.\ Bachmann, K.\ Goede, A.\ G.\ Beck-Sickinger, M.\
  Grundmann, A.\ Irb\"ack, and W.\ Janke, Angew.\ Chem.\ Int.\ Ed.\
  \href{http://onlinelibrary.wiley.com/doi/10.1002/anie.201000984/abstract}{\textbf{49},
  9530 (2010)}.
%
\bibitem{mb1}
M.~Bachmann, \emph{Thermodynamics and Statistical Mechanics of
Macromolecular Systems}, (Cambridge University Press, Cambridge, 2014).
%
\bibitem{vbj1} T.\ Vogel, M.\ Bachmann, and W.\ Janke, Phys.\ Rev.\ E
  \href{http://dx.doi.org/10.1103/PhysRevE.76.061803}{\textbf{76},
  061803} (2007).
%
\bibitem{svbj1}
S.\ Schnabel, T.\ Vogel, M.\ Bachmann, and W.\ Janke, Chem.\ Phys.\ Lett.\
\href{http://dx.doi.org/10.1016/j.cplett.2009.05.052}{\textbf{476}, 201} (2009).
%
\bibitem{sbj3} S.\ Schnabel, M.\ Bachmann, and W.\ Janke, J.\ Chem.\
  Phys.\ \href{http://dx.doi.org/10.1063/1.3223720}{\textbf{131},
    124904} (2009).
%
\bibitem{bogner1}
T.\ Bogner, A.\ Degenhard, and F.\ Schmid, Phys.\ Rev.\ Lett.\ \textbf{93},
268108 (2004).
%
\bibitem{allen3}
A.~Swetnam and M.~P.~Allen, Phys.\ Rev.\ E \textbf{85}, 062901 (2012).
%
\bibitem{mjb2014} M.~M\"oddel, W.~Janke, and M.~Bachmann, Phys.\ Rev.\
  Lett.\
  \href{http://dx.doi.org/10.1103/PhysRevLett.112.148303}{\textbf{112},
    148303} (2014).
%
\bibitem{vrbova}
T.\ Vrbov\'{a} and S.\ G.\ Whittington, J.\ Phys.\ A \textbf{29}, 6253
(1996); J.\ Phys.\ A \textbf{31}, 3989 (1998);
T.\ Vrbov\'{a} and K.\ Proch\'{a}zka, J.\ Phys.\ A \textbf{32}, 5469
(1999).
%
\bibitem{singh}
Y.\ Singh, D.\ Giri, and S.\ Kumar, J.\ Phys.\ A \textbf{34}, L67
(2001); R.\ Rajesh, D.\ Dhar, D.\ Giri, S.\ Kumar, and Y.\ Singh, Phys.\ 
Rev.~E \textbf{65}, 056124 (2002).
%
\bibitem{causo1}
M.\ S.\ Causo, J.\ Chem.\ Phys.\ \textbf{117}, 6789 (2002).
%
\bibitem{prellberg1}
J.\ Krawczyk, T.\ Prellberg, A.\ L.\ Owczarek, and A.\ Rechnitzer,
Europhys.\ Lett.\ \textbf{70}, 726 (2005).
%
\bibitem{bj4} M.\ Bachmann and W.\ Janke, Phys.\ Rev.\ Lett.\
  \href{http://dx.doi.org/10.1103/PhysRevLett.95.058102}{\textbf{95},
    058102} (2005); Phys.\ Rev.\ E
  \href{http://dx.doi.org/10.1103/PhysRevE.73.041802}{\textbf{73},
    041802} (2006); Phys.\ Rev.\ E
  \href{http://dx.doi.org/10.1103/PhysRevE.73.020901}{\textbf{73},
    020901(R)} (2006).
%
\bibitem{linse1}
N.\ K\"allrot and P.\ Linse, Macromolecules \textbf{40}, 4669 (2007).
%
\bibitem{paul1}
J.\ Luettmer-Strathmann, F.\ Rampf, W.\ Paul, and K.\ Binder, J.\ Chem.\
Phys.\ \textbf{128}, 064903 (2008).
%
\bibitem{mbj1} M.\ M\"oddel, M.\ Bachmann, and W.\ Janke, J.\ Phys.\
  Chem.\ B
  \href{http://pubs.acs.org/doi/abs/10.1021/jp808124v}{\textbf{113},
    3314} (2009).
%
\bibitem{liang1}
L.~Wang, T.~Chen, X.~Lin, Y.~Liu, H.~Liang, J.\ Chem.\ Phys.\ \textbf{131},
244902 (2009).
%
\bibitem{allen1}
A.\ D.\ Swetnam and M.\ P.\ Allen, Phys.\ Chem.\ Chem.\ Phys.\ \textbf{11},
2046 (2009).
%
\bibitem{ivanov09jpcb}%
V.~A.\ Ivanov, J.~A.\ Martemyanova, M.~M\"uller, W.~Paul, and
K.~Binder, J.~Phys.\ Chem.\ B \textbf{113}, 3653 (2009).
%
\bibitem{mjb2} M.~M\"oddel, W.~Janke, and M.~Bachmann, Phys.\ Chem.\
  Chem.\ Phys.\
  \href{http://pubs.rsc.org/en/content/articlelanding/2010/cp/c002862b}{\textbf{12},
    11548} (2010); Macromolecules
  \href{http://pubs.acs.org/doi/abs/10.1021/ma201307c}{\textbf{44},
    9013} (2011).
%
\bibitem{ywli1}
Y.~W.~Li, T.~W\"ust, and D.~P.~Landau, Phys.\ Rev.~E \textbf{87}, 012706
(2013).
%
\bibitem{popova}
H.~Popova and A.~Milchev, J.~Chem.\ Phys.\
\textbf{127}, 194903 (2007); \textit{ibid.} \textbf{129},
215103 (2008).
%
\bibitem{kjb1} S.\ Karalus, W.\ Janke, and M.\ Bachmann, Phys.\ Rev.\
  E \href{http://dx.doi.org/10.1103/PhysRevE.84.031803}{\textbf{84},
    031803} (2011).
%
\bibitem{milchev1}
A.~Milchev and K.~Binder, J.~Chem.\ Phys.\ \textbf{117}, 6852
(2002).
%
\bibitem{srebnik}
I.~Gurevitch and S.~Srebnik, Chem.\ Phys.\ Lett.\ \textbf{444}, 96 (2007);
J.~Chem.\ Phys.\ \textbf{128}, 144901 (2008).
%
\bibitem{gvb1}
J.~Gross, T.~Vogel, and M.~Bachmann, \textit{Structural 
Phases of Adsorption for Flexible Polymers on Nanocylinder Surfaces}, to be published (2015).
%
\bibitem{dressel01tap}%
M.~S.\ Dresselhaus, G.~Dresselhaus, and P.~Avouris, eds.,
\textit{Carbon Nanotubes: Synthesis, Structure, Properties, and
Applications}, Topics in Applied Physics, Vol.~\textbf{80} (Springer,
Heidelberg, 2001).
%
\bibitem{jorio08buch}%
A.~Jorio, G.~Dresselhaus, and M.~S.\ Dresselhaus, eds.,
\textit{Carbon Nanotubes: Advanced Topics in the Synthesis,
Structure, Properties, and Applications}, Topics in Applied
Physics, Vol.~\textbf{111} (Springer, Berlin, Heidelberg, 2008).
%
\bibitem{hasan09am}%
T.~Hasan, Z.~Sun, F.~Wang, F.~Bonaccorso, P.~H.\ Tan, A.~G.\ Rozhin,
and A.~C.\ Ferrari, Adv. Mater.\ \textbf{21}, 3874 (2009).
%
\bibitem{gao03ea}%
M.~Gao, L.~Dai, and G.~G.\ Wallace, Electroanalysis \textbf{15}, 1089
(2003).
%
\bibitem{wang10nanolett}%
L.~Wang, Y.~Liu, W.~Li, X.~Jiang, Y.~Ji, X.~Wu, L.~Xu, Y.~Qiu,
K.~Zhao, T.~Wei, Y.~Li, Y.~Zhao, and C.~Chen, Nano Lett.\
\textbf{11}, 772 (2011).\vadjust{\break}
%
\bibitem{brewer05langmuir}%
S.~H.\ Brewer, W.~R.\ Glomm, M.~C.\ Johnson, M.~K.\ Knag, and
S.~Franzen, Langmuir \textbf{21}, 9303 (2005).
%
\bibitem{tran08nanolett}%
M.~Q.\ Tran, J.~T.\ Cabral, M.~S.~P.\ Shaffer, and A.~Bismarck, Nano
Lett.\ \textbf{8}, 2744 (2008).
%
\bibitem{wei06nanolett}%
C.~Wei, Nano Lett.\ \textbf{6}, 1627 (2006).
%
\bibitem{tallury10jpcb}%
S.~S.\ Tallury and M.~A.\ Pasquinelli, J.~Phys.\ Chem.~B \textbf{114},
4122 (2010); \textit{ibid.}\ 9349 (2010).
%
\bibitem{liu08jpcc}%
W.~Liu, C.-L.\ Yang, Y.-T.\ Zhu, and M.-S.\ Wang, J.~Phys.\ Chem.~C
\textbf{112}, 1803 (2008).
%
\bibitem{caddeo10jpcc}%
C.~Caddeo, C.~Melis, L.~Colombo, and A.~Mattoni, J.~Phys. Chem.~C
\textbf{114}, 21109 (2010).
%
\bibitem{wallace07nanolett}%
E.~J.\ Wallace and M.~S.~P.\ Sansom, Nano Lett.\ \textbf{7}, 1923
(2007).
%
\bibitem{wallace10scale}%
E.~J.\ Wallace, R.~S.~G.\ D'Rozario, B.~M.\ Sanchez, and M.~S.~P.\
Sansom, Nanoscale \textbf{2}, 967 (2010).
%
\bibitem{angelik10lang}%
P.~Angelikopoulos and H.~Bock, Langmuir \textbf{26}, 899 (2010).
%
\bibitem{vb1}%
  T.~Vogel and M.~Bachmann, Phys.\ Rev.\ Lett.\
  \href{http://dx.doi.org/10.1103/PhysRevLett.104.198302}{\textbf{104},
    198302} (2010); Phys.\ Proc.\
  \href{http://dx.doi.org/10.1016/j.phpro.2010.08.020}{\textbf{4},
    161} (2010)
%
\bibitem{wagner90jap}%
H.~D.\ Wagner, J.~Appl.\ Phys.\ \textbf{67}, 1352 (1990).
%
\bibitem{carroll86langmuir}%
B.~J.\ Carroll, Langmuir \textbf{2}, 248 (1986).
%
\bibitem{glaeser79jur}%
R.~Glaeser, W.~Chiu, and D.~Grano, J.~Ultrastruct.\ Res.\ \textbf{66},
235 (1979).
%
\bibitem{infl1} S.~Schnabel, D.~T.\ Seaton, D.~P.\ Landau, and
  M.~Bachmann, Phys.\ Rev.~E,
  \href{http://dx.doi.org/10.1103/PhysRevE.84.011127}{\textbf{84},
    011127} (2011).
%
\bibitem{still1}
F.~H.\ Stillinger, T.~Head-Gordon, and C.~L.\ Hirshfeld, Phys.\ Rev.~E
\textbf{48}, 1469 (1993);
F.~H.\ Stillinger and T.\ Head-Gordon, Phys.\ Rev.~E \textbf{52}, 2872
(1995).
%
\bibitem{baj1} M.~Bachmann, H.~Ark{\i}n, and W.~Janke, Phys.\ Rev.~E
  \href{http://dx.doi.org/10.1103/PhysRevE.71.031906}{\textbf{71},
    031906} (2005).
%
\bibitem{jbj1} C.~Junghans, M.~Bachmann, and W.~Janke, Phys.\ Rev.\
  Lett.\
  \href{http://dx.doi.org/10.1103/PhysRevLett.97.218103}{\textbf{97},
    218103} (2006); J.~Chem.\ Phys.\
  \href{http://dx.doi.org/10.1063/1.2830233}{\textbf{128}, 085103}
  (2008).
%
\bibitem{vb2}%
  T.~Vogel and M.~Bachmann, Comp.\ Phys.\ Comm.\
  \href{http://dx.doi.org/10.1016/j.cpc.2010.11.007}{\textbf{182},
    1928} (2011).
%
\bibitem{muca} 
B.~A.\ Berg and T.~Neuhaus, Phys.\ Lett.~B \textbf{267}, 249 (1991); Phys.\
Rev.\ Lett.\ \textbf{68}, 9 (1992).
%
\bibitem{janke98physA} 
W.~Janke, Physica A \textbf{254}, 164 (1998).
%
\bibitem{mb2} M.~Bachmann, Phys.\ Scr.\
  \href{http://dx.doi.org/10.1088/0031-8949/87/05/058504}{\textbf{87},
  058504} (2013).
%
\bibitem{wl_prle} 
F.~Wang and D.~P.\ Landau, Phys.\ Rev.\ Lett.\
\textbf{86}, 2050 (2001); Phys.\ Rev.~E \textbf{64}, 056101 (2001).
%
\bibitem{zhou05pre} 
C.~Zhou and R.~N.\ Bhatt, Phys.\ Rev.~E \textbf{72}, 025701(R) (2005).
%
\bibitem{vmab1} T.~Vogel, T.~Mutat, J.~Adler, and M.~Bachmann,
  Commun.\ Comp.\ Phys.\
  \href{http://www.global-sci.com/freedownload/v13_1245.pdf}{\textbf{13},
    1245} (2013); Phys.\ Proc.\
  \href{http://dx.doi.org/10.1016/j.phpro.2011.06.005}{\textbf{15},
    87} (2011).
%
\bibitem{wang1}
For a review, see C.~Shen, A.~H.~Brozena, Y.~Wang, Nanoscale \textbf{3}, 503
(2011). 
%
\bibitem{bj1} M.~Bachmann and W.~Janke, J.~Chem.~Phys.\
  \href{http://dx.doi.org/10.1063/1.1651055}{\textbf{120}, 6779} (2004).
%
\bibitem{gross1}
D.~H.~E.~Gross, \emph{Microcanonical Thermodynamics} (World Scientific,
Singapore, 2001).
%
\bibitem{kremer1}
E.~Eisenriegler, K.~Kremer, and K.~Binder, J.~Chem.\ Phys.\ \textbf{77}, 6296
(1982).\vadjust{\break}
%
\end{thebibliography}
\end{document}